\documentclass{ws-procs9x6}

\begin{document}

\title{Charm Production at Large Rapidities in p+p and d+Au Collisions at PHENIX at RHIC}

\author{Xiaorong Wang for PHENIX collaboration}

\address{Physics Department, New Mexico State University, Las Cruces, NM88003, USA\\
Hua-Zhong Normal University, Wuhan 430079, P.R. China\\
E-mail: xrwang@bnl.gov}

\maketitle

\abstracts{We study charm production through dimuon and single
muon measurements at forward and backward rapidities in p+p and
d+Au collisions with the PHENIX muon detectors. We also compare
open charm to $J/\psi$ yields in the forward and backward
rapidities in d+Au collisions and study the origin of the large
forward and backward asymmetry in open charm production observed
by the PHENIX experiment. }

\section{Introduction: }

Charm quarks are believed to be mostly created from initial gluon
fusion in hadronic collisions. Since they are massive, heavy
flavor hadrons are proposed to be ideal probes to study the early
stage dynamics in heavy-ion collisions. Measurements of heavy
quark production in p+p collisions serve as important tests for
perturbative Quantum Chromo Dynamics (pQCD).

$J/\psi$ and open charm productions are two of the most important
hard probes of the hot dense matter created in Au+Au collisions.
It is necessary to understand their production in the cold nuclear
medium in d+Au collisions as a reference in their production in
Au+Au collisions.

Since the initial formation of both open and closed charm is
sensitive to initial gluon densities, then gluon structure
functions, shadowing or anti-shadowing and initial state energy
loss will all effect their production. During the hadronization,
the $J/\psi$ production mechanism is different from that of open
charm. In the final state, the $J/\psi$ can be disassociated or
absorbed, but for open charm, the main nuclear medium effect is
final state multiple scattering and energy loss. The comparison of
open charm and the $J/\psi$ production will help us to understand
$J/\psi$ production mechanism and disentangle different initial
state and final state nuclear medium effects.

In $\sqrt{s_{NN}} = 200$ GeV d+Au collisions at RHIC, measurements
at forward rapidity (deuteron direction) probe the shadowing
region while the anti-shadowing region is probed in backward
rapidity. Recent models of gluon shadowing\cite{glouonshadowing},
color glass condensate\cite{CGC} and
recombination\cite{recombination} are implemented to understand
open charm production at forward rapidity.

\section{Charm measurement at PHENIX}
The PHENIX experiment\cite{phenix} has measured $J/\psi$ and open
charm production through observation of dilepton and semi-leptonic
decays at forward and backward rapidity with the PHENIX muon
spectrometers, covering both forward and backward directions in
the rapidity range of $1.2 < |\eta| < 2.4$.

\section{Open Charm results in p+p and dAu collisions}
A PYTHIA simulation shows around 75$\%$ of prompt muons with $p_T
>$ 0.9 GeV/$c$ come from open charm decay in d+Au collisions, while 11$\%$
come from open bottom decay. Prompt muons are produced close to
the collision vertex. We can separate heavy flavor decays and
light hadron decays experimentally by studying the shape of the
vertex distribution.

\begin{figure}[ht]
%\epsfxsize=10cm   %width of figure - will enlarge/reduce the figures
%\epsfbox{fig3.eps}
%\figurebox{2cm}{3cm}{} %to have a box alone
\centerline{\epsfxsize=4.1in\epsfbox{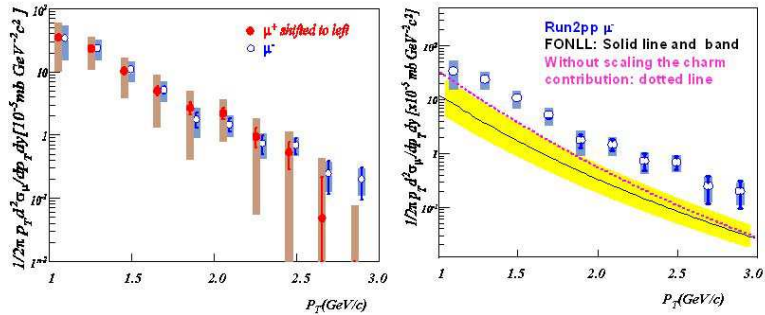}}
\vskip -2cm \caption{Left: $p_T$ spectrum of prompt muons. Error
bars indicate statistical errors and shaded bands indicate
systematic errors. Right: The measured $p_T$ spectrum of negative
prompt muons, the PYTHIA prediction without scaling the charm
contribution (dotted line), and a FONLL calculation (solid line
with systematic error band). \label{inter}}
\end{figure}

The invariant differential cross section for muon candidate
production at forward rapidity ($1.5<\eta<1.8 $) has been measured
by PHENIX over the transverse momentum range $1<p_T<3$ GeV/$c$ in
200 GeV p+p collisions\cite{run2pp}. The resulting muon spectrum
from heavy flavor decays is compared to PYTHIA and a
next-to-leading order perturbative QCD calculation, shown in
Figure 1. PHENIX muon arm data (at forward and backward rapidity)
is compatible with the PHENIX charm measurement at y = 0
\cite{singleepp}, and it exceeds predictions from PYTHIA and
FONLL.

The nuclear modification factor of d+Au collisions is defined as
the particle yield per nucleon-nucleon collision relative to the
yield in p+p collisions. The nuclear modification factors for
prompt muons are shown in Figure 2. Prompt muon production shows
suppression at forward rapidity and enhancement in the backward
direction.
\begin{figure}[ht]
%\epsfxsize=10cm  %width of figure - will enlarge/reduce the figures
%\epsfbox{fig3.eps}
%\figurebox{2cm}{3cm}{} %to have a box alone
\centerline{\epsfxsize=4.1in\epsfbox{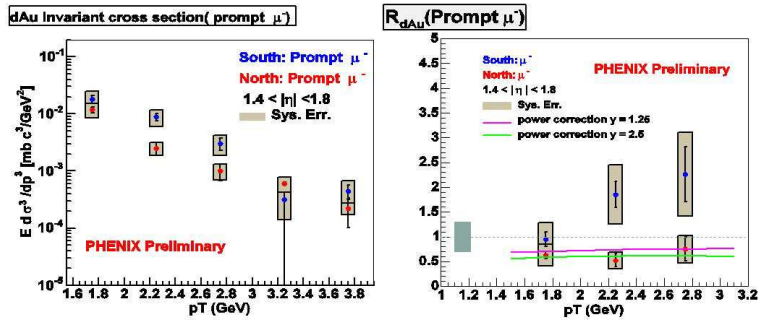}} \vskip
-1cm
 \caption{Invariant spectra of the prompt muon (left) and
nuclear modification factor of prompt muons(right) in d+Au
collisions. The theoretical curves are come from power correction
model at $\eta = 1.25$ and 2.5. \label{inter}}
\end{figure}

\section{$J/\psi$ results in p+p and dAu collisions}

PHENIX measured $J/\psi$ production in forward, backward and
central rapidity in p+p and d+Au collisions at 200 GeV
\cite{jpsi038}.

Figure 3a shows the measured pp differential cross section times
branching ratio vs rapidity. A fit to a shape generated with
PYTHIA is performed, and using a di-lepton branching ratio of
$5.9\%$ gives a total cross section $\sigma_{pp}^{J/\psi} =
2.61\pm 0.20(fit) \pm 0.26(abs) \mu$b. Figure 3b shows the nuclear
modification factor vs rapidity. It is significantly lower at the
forward rapidity. Theoretical calculations that include the
effects of absorption and shadowing are shown in the Figure 3b.
The data favor a modest shadowing rather than the stronger gluon
shadowing.

\begin{figure}[ht]
%\epsfxsize=10cm   %width of figure - will enlarge/reduce the figures
%\epsfbox{fig3.eps}
%\figurebox{2cm}{3cm}{} %to have a box alone
\centerline{\epsfxsize=3.6in\epsfbox{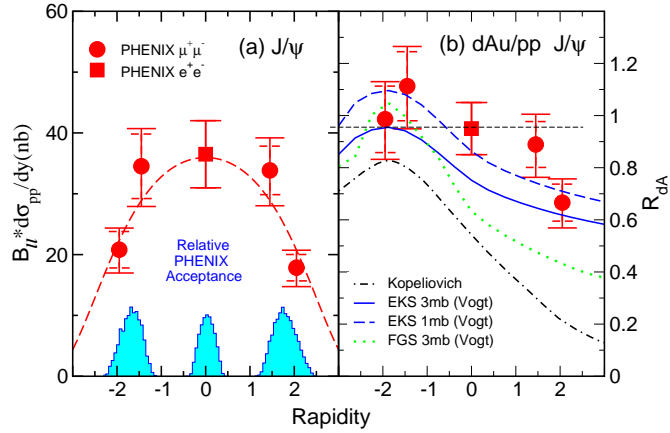}} \vskip -0.5cm
\caption{(a) $J/\psi$ differential cross section times di-lepton
branching ratio vs rapidity. (b) The minibias $R_{dAu}$ vs
rapidity. \label{inter}}
\end{figure}

\section{Summary and outlook}
We observe a significant cold nuclear medium effect in charm
production in forward and backward rapidity in d+Au collisions at
200 GeV/$c$. Both open charm and $J/\psi$ results show a
suppression in forward rapidity. The open charm data are
consistent with CGC and power correction model\cite{ivan}; the
$J/\psi$ data favor shadowing with weak absorption. At backward
rapidity, open charm results show enhancement while $J/\psi$
results are consistent with unity.

We need a more precise d+Au measurements and more theoretical work
to understand the cold nuclear medium effects as a baseline for
understanding the hot dense matter produced in Au+Au collisions.

%\section*{}

\end{document}